# Development of Mobile-Interfaced Machine Learning-Based Predictive Models for Improving Students' Performance in Programming Courses

Fagbola Temitayo Matthew
Department of Computer Science
Federal University, Oye-Ekiti, Nigeria

Obe Olumide
Department of Computer Science
Federal University of Technology, Akure

Adeyanju Ibrahim Adepoju
Department of Computer Engineering
Federal University, Oye-Ekiti, Nigeria

Olaniyan Olatayo, Esan Adebimpe, Omodunbi Bolaji
Department of Computer Engineering
Federal University, Oye-Ekiti, Nigeria

Oloyede Ayodele
Department of Computer Science
Caleb University, Imota, Lagos, Nigeria

Egbetola Funmilola
Department of Computer Science & Engineering
LAUTECH, Ogbomoso, Nigeria

*Abstract*—Student performance modelling (SPM) is a critical step to assessing and improving students' performances in their learning discourse. However, most existing SPM are based on statistical approaches, which on one hand are based on probability, depicting that results are based on estimation; and on the other hand, actual influences of hidden factors that are peculiar to students, lecturers, learning environment and the family, together with their overall effect on student performance have not been exhaustively investigated. In this paper, Student Performance Models (SPM) for improving students' performance in programming courses were developed using M5P Decision Tree (MDT) and Linear Regression Classifier (LRC). The data used was gathered using a structured questionnaire from 295 students in 200 and 300 levels of study who offered Web programming, C or JAVA at Federal University, Oye-Ekiti, Nigeria between 2012 and 2016. Hidden factors that are significant to students' performance in programming were identified. The relevant data gathered, normalized, coded and prepared as variable and factor datasets, and fed into the MDT algorithm and LRC to develop the predictive models. The developed models were obtained, validated and afterwards implemented in an Android 1.0.1 Studio environment. Extended Markup Language (XML) and Java were used for the design of the Graphical User Interface (GUI) and the logical implementation of the developed models as a mobile calculator, respectively. However, Mean Absolute Error (MAE), Root Mean Squared Error (RMSE), Relative Absolute Error (RAE) and the Root Relative Squared Error (RRSE) were the metrics used to evaluate the robustness of MDT and LRC models. The evaluation results obtained indicate that the variable-based LRC produced the best model in terms of MAE, RMSE, RAE and the RRSE having yielded the least values in all the evaluations conducted. Further results obtained established the strong significance of attitude of students and lecturers, fearful perception of students, erratic power supply, university facilities, student health and students' attendance to the performance of students in programming courses. The variable-based LRC model presented in this paper could provide baseline information about students' performance thereby offering better decision making towards improving teaching/learning outcomes in programming courses.

## I. INTRODUCTION

Computer programming courses are a fundamental part of many Universities' curricula and among the most important subjects for computer science and information technology students. This requires the knowledge of programming tools and languages, problem-solving skills and effective strategies for program design and implementation [1]. Furthermore, students are being exposed to various programming specifications and techniques which normally entails an overview of algorithms, concept of programming, basic data structure, problem analysis and illustrations describing the application of various techniques to problems which are quite difficult to understand [2]. Furthermore, the high level of abstraction and very complex language syntax and semantic structures induced in programming makes it a much dreaded task in which most students fail [2]. This is evidenced by the notion that the same set of students who failed programming courses performed better in other non-programming courses [3]. As a matter of fact, the failure rate in programming courses at the University level suggests that learning to program is a difficult task [3]. The perception of the complexity ascribed to programming courses can be described as one of the main reasons that may have attributed to the decline in number of undergraduates who offer or intend to offer computer science in various institutions [4].

Chermahini [5] noted that students are different based on their ability to learn, how they respond to instructional practices, their motivational differences from one individual to another and that the more students understand the differences in their abilities, the better are the chances they have to meet their different learning needs in order to achieve good scores in examinations. Students' performance is majorly affected by several social, economic, institutional, environmental, psychological and personal factors which vary across individuals and regions [6]-[8]. Unfortunately, poor performances have ravaged the academic institutions due to indices of those factors which influence students' performance including poor funding, lack of frequent curricular review, overpopulation, students' unrest, staff strikes, poor facilities, coarse relations between the university and government, inadequate teaching and research facilities needed to enhance students' learning and performance. More specifically, Ogbogu [6] and Irfan and Shabana [9] emphasized that challenges such as poorly equipped departmental and central libraries, overcrowded lecture rooms, method of collating and accessing semester results, interruption of electricity supply, poor access to internet facilities, incessant strike and closure of school and poor accommodation facilities which are pertinent to developing countries affect student performance.

Students' performance assessment has become a pressing issue that requires fair attention from all regardless of differences in interest and intentions [9], [10]. However, different methods have been used to evaluate students' performance, and more than ever before, information generated by evaluation can be helpful for students and tutors to take timely, meaningful and effective decisions. Most existing student performance models have adopted statistical techniques for prediction which are probability-induced, depicting that results may not be scientifically correct but rather are based on estimation. To this end, several authors have adopted data mining and soft computing techniques in educational domain and/or to evaluate students' performance [11]-[17].

Ashish, Saeed, Maizatul, and Hamidreza [14] focused on consolidating the different types of clustering algorithms been applied within the context of Educational Data Mining (EDM) to harnessing the power of the massive didactic data recently being generated in institutions. EDM was employed to analyze data generated in an educational setup by the various intra-connected systems in a bid to develop a model for improving learning and institutional effectiveness. Among the slightly numerous clustering algorithm consolidated by the authors are Expectation Maximization, Hierarchical Clustering, Simple *k*-Means and *x*-Means, Apriori Algorithm (as applied to academic records of students in a guise to obtain the best association rules which helps in student profiling), *C*-Means clustering, Ward's clustering, Markov Clustering (MCL) algorithm, Unique Clustering with Affinity Measure (UCAM), Fuzzy sets, Transitive Closure and a hierarchical cluster analysis which was performed on the questionnaire data. As concluded by these authors, data mining methods in the educational sector sets to uncover the previously hidden data to meaningful information that can be used for strategic and learning gains.

Kolo, Adepoju and Alhassan [18] aimed at predicting the performance of students with the decision tree approach. Gurmeet and Williamjit [13] employed data-mining approach for an effective prediction of student performance based on personal, social, psychological and environmental variables. This was to ensure a high accuracy in the prediction of student performance, thereby assisting to identify students with low academic achievements. The parameters employed in the study include gender, hometown, family income, previous semester grade, attendance, communication language (medium), seminar performance and participation in sports. Analysis of these parameters was conducted by implementing the algorithms in WEKA tool. Naïve Bayes and J48 algorithms were used for classification and the result showed that the Naive Bayes algorithm provided an accuracy of 63.59% while the J48 algorithm provided an accuracy of 61.53%.

Generally, the educational sector in developing countries is being faced by a series of multi-factored challenges that contribute to the rapid decline in the performance of students located within such contemporary environments. Teachers and students alike have for so long been unable to estimate the impact that certain factors have on academic performances but rather anticipate good performances in the long run. This way, it becomes impossible for student to quickly re-adjust and retune performance demeaning challenges surrounding them or probably their responses to such surrounding factors. More often than not, the actual influences of hidden factors that are peculiar to students, lecturers, learning environment and the family, together with their overall effect on student performance have not been exhaustively investigated in existing studies.

In this paper, M5P decision tree and linear regression classifier, which are among the most widely adopted machine learning techniques, are employed to develop the student performance predictive models. Metrics used to evaluate the performance of the machine learning techniques employed include mean absolute error, root mean squared error, relative absolute error and the root relative squared error, correlation coefficient, time taken to build the model and the time taken to test the model.

The major contributions of this paper are as follows:

*a) Exhaustively investigated, examined, identified and established new hidden factors and associated variables on which students' performance in programming courses is dependent and that are particularly peculiar to a prototype University in a developing economy. These are significant and technical extensions beyond most student performance models that currently exist;*

*b) Beyond the spheres of statistical approaches commonly used for student performance modeling which are based on probability and estimation in most existing works, this study applied machine learning techniques (M5P Decision Tree and Linear Regression Classifier) to predicting*

student performance in programming courses to guarantee precision and accuracy of the resultant predictive models;

*c) Towards facilitating the accessibility, availability and ubiquity of the developed predictive models, a mobile application, that visually interfaces the stakeholders and all student performance indices with the models, was developed. This is to realize real-time use in predicting students' performance and for promoting effective and efficient decision making on education planning by all stakeholders.*

The rest of this paper is organized as follows: Section 2 discusses the materials and method including the M5P decision tree and linear regression classifier, data acquisition, the development and validation of the machine learning-based predictive models and the performance evaluation metrics for the machine-learning based approaches. In Section 3, the design and implementation of the mobile-frontend application for the developed predictive models are presented and discussed. The results of performance evaluation of the machine learning approaches are presented and discussed in Section 4 while the conclusion and future works are presented in Section 5.

## II. MATERIALS AND METHOD

In this research, models for predicting students' performance in programming courses were developed based on M5P and linear regression classification algorithms in three basic steps. These include data acquisition, development of the predictive models and finally model validation. Furthermore, the performance evaluation of the machine learning approaches employed and the mobile implementation of the predictive models developed were conducted.

### A. The Classification Algorithms

*1) M5P Decision Tree:* This is a decision tree model that learns regression tasks. The M5P learns efficiently and can cope with highly-dimensional data with up to several hundreds of distinct attributes. According to Quinlan [19], M5P decision tree is the most accurate among the family of regression tree learners with much smaller model trees than regression trees. It uses mean squared error as the impurity function. A M5P tree is constructed by recursive partitioning of a data into a collection of set T which can either be associated with a leaf or a split function that segregates $T$ into some subsets based on some split function criteria [20]. The subsets that emerge are further partitioned following the same process repeatedly. However, the quality of split (goodness of fit) is evaluated using a function $\emptyset(S, T)$ where $S$ is the split candidate in node $T$ such that the split candidate that maximizes the value of quality of fit is selected as the next node of tree [21]. That is,

$$\emptyset(S, T) = \Delta I(S, T) = I(T) - {}_{PL}I(T_L) - {}_{PR}I(T_R) \qquad (1)$$

where $I(T)$ is the impurity function at node $T$ for $k$ classes in a dataset defined as:

$$I(T) = I(p(1|T), p(2|T), \dots, p(k|T)) \qquad (2)$$

$PL$ and $PR$ are the probabilities that an instance is going to the left branch and right branch of $T$ according to split $S$, $p(j|T)$ is the estimated posterior probability of class $j$ given a point in node $T$, $\Delta I(S, T)$ is the difference between the impunity measure of node $T$ and two child nodes $T_L$, $T_R$ according to split $S$. The information gain in M5P is determined by the difference in the values of standard deviation obtained before and after the split function test. Simply put, given data $T$, where $T_i$ denotes the subsets of $T$ corresponding to the $i_{th}$ outcome of a split function test, then the expected error reduction value is determined by Hieu [22]:

$$\Delta error = sd(T) - \sum_i \frac{|T_i|}{|T|} sd(T_i) \qquad (3)$$

The split function test criterion that maximizes this expected error reduction is then selected. To avoid overfitting, subtrees that do not improve the performance of the tree are pruned via an error-based estimation procedure, from the leaves to the root node [23]. This is determined by the difference in the estimated error of a node and estimated error of the subtree below at each internal node.

*2) Linear Regression Classifier:* The linear regression classifier is a mathematical measure depicting the mean relationship among two or more variables based on the original units of the data [24]. This often involves the estimation and prediction of an unknown value of one variable from the known value of another variable [25]. This implies that there exists a linear regression between the variables should the regression curve be a straight line. With linear regression, the values of the dependent variable increase by a constant absolute amount for a unit change in the value of the independent variable. However, the general form of linear regression measure is given as [26]:

$$h_\theta(x) = \theta_0 x_0 + \theta_1 x_1 + \theta_2 x_2 + \theta_3 x_3 + \dots + \theta_n x_n = \theta^T(x) \qquad (4)$$

where $h_\theta(x)$ *is a straight line with variable intercept*

*and parameters* $\theta_0, \theta_1, \dots, \theta_n$, if $x_0 = 0$ is assumed.

**Algorithm**: Linear Regression Classification [27]
**Inputs**: Class models $X_i \in R^{q \times pi}$, $i = 1, 2, \dots, N$ and a test input student performance factors' vector $y \in R^{q \times 1}$.
**Output**: Class of $y$

   i. For each class model, $\hat{\beta}_i \in R^{pi \times 1}$ is evaluated such that $\hat{\beta}_i = (X_i^T X_i)^{-1} X_i^T y$, $i = 1, 2, \dots, N$

   ii. $\hat{y}_i$ is computed for each $\hat{\beta}_i$, $\hat{y}_i = X_i \hat{\beta}_i$, $i = 1, 2, \dots, N$

   iii. Distance between original and predicted response variables is determined by $d_i(y) = ||y - \hat{y}_i||_{2}$, $i = 1, 2, \dots, N$

   iv. Decision is made with regard to the class that has the minimum distance $d_i(y)$

### B. Data Acquisition

Hidden factors that are significant to student performance were identified via a thorough literature review, interview and field observations. Questionnaire was developed for the University under study with respect to information on programming courses and associated scores as presented at the Appendix section. In Table I, the contextual definition of the variables is presented. Copies of the questionnaires were disseminated to students that had offered programming courses and their respective lecturers in the University.

Relevant data were gathered, normalized and coded. The coded data was utilized by the machine learning techniques to develop the student performance models and were further validated for prediction purpose.

TABLE I.        VARIABLE EXPRESSION FROM DESIGNED QUESTIONNAIRE

| S/N | Expressions |
|---|---|
| $x_1$ | I had enough time to study programming |
| $x_2$ | Studying before attending a class aided my assimilation during programming classes. |
| $x_3$ | Studying programming was never a wasted effort |
| $x_4$ | Programming sounded very scary |
| $x_5$ | I was always nervous during programming classes |
| $x_6$ | I was always nervous during programming examinations |
| $x_7$ | I attended programming classes regularly |
| $x_8$ | Blending in after missing a class was very easy |
| $x_9$ | I was very serious with programming classes |
| $x_{10}$ | I believed I could understand the programming course |
| $x_{11}$ | I had interest in programming beyond class level |
| $x_{12}$ | Programming was not confusing and did not cause headache |
| $x_{13}$ | Programming is relevant to my pursuit |
| $x_{14}$ | Group discussions helped me to understand programming |
| $x_{15}$ | Attending programming tutorials was very helpful |
| $x_{16}$ | Programming courses tutorials helped me so much |
| $x_{17}$ | Motivation of programming lecturers encouraged my commitment towards learning programming |
| $x_{18}$ | Programming language lecturers helped me develop interest in programming |
| $x_{19}$ | Programming languages lecturers were never partial in their dealings with students |
| $x_{20}$ | Programming lecturers were friendly during lectures |
| $x_{21}$ | Programming language lecturers enforced discipline during their lectures |
| $x_{22}$ | Programming languages lecturers were too serious during lectures |
| $x_{23}$ | Teaching methods and styles of programming lecturers inhibited lecture clarity |
| $x_{24}$ | Programming language lecturers wasted time on matters with less relevance in class |
| $x_{25}$ | Programming language lecturers were always clear, precise and communicates understandably |
| $x_{26}$ | Programming language lecturers made use of enough relevant instructional materials |
| $x_{27}$ | Programming language lecturers delivered course contents well and to my understanding |
| $x_{28}$ | Programming language lecturers were very clear and explicit |
| $x_{29}$ | Programming language lecturers didn't miss classes |
| $x_{30}$ | Programming language lecturers attended to me whenever I had difficulties with their course(s) |
| $x_{31}$ | Programming lecturers were always available |
| $x_{32}$ | Programming course lecturers allowed students to ask questions and take time to explain |
| $x_{33}$ | Programming course lecturers came to class fully prepared |
| $x_{34}$ | Programming languages lecturers spent extra time to explain things during class |
| $x_{35}$ | Programming language lecturers usually came early to class |
| $x_{36}$ | I fell sick quite often |
| $x_{37}$ | Prolong usage of computer caused me headache |
| $x_{38}$ | I took a few compulsory medications frequently |
| $x_{39}$ | It was difficult to charge my computer even within the campus |
| $x_{40}$ | Erratic power supply reduced the effectiveness of my practice |
| $x_{41}$ | Consistent power supply helped me in programming courses |
| $x_{42}$ | I had a good background in physics |
| $x_{43}$ | I had a good background in mathematics |
| $x_{44}$ | I had a good background in English |
| $x_{45}$ | Strong background in Physics and Mathematics helped me in programming |
| $x_{46}$ | Absence of accessible ICT facilities inhibited my programming performance |
| $x_{47}$ | The environment where we had programming lectures was not conducive |
| $x_{48}$ | Lack of computer programming facilities disrupted clear understanding of programming lessons |
| $x_{49}$ | The school library was not equipped with materials relevant to programming |
| $x_{50}$ | Large class population disrupted my concentration during programming lectures |
| $x_{51}$ | Population of students offering programming courses debarred my commitment to learning |
| $x_{52}$ | Effectiveness of the programming lecturers' teaching was reduced by huge programming class population. |
| $x_{53}$ | Programming lectures were scheduled after an equally tiring lecture |
| $x_{54}$ | Programming courses were scheduled to non-conducive times |
| $x_{55}$ | We had programming classes at unfavorable times |
| $x_{56}$ | Programming lecture theatres were equipped with audio-visuals and learning aids |
| $x_{57}$ | Programming courses were analyzed clearly to sight |
| $x_{58}$ | I had a visual understanding of what the programming lecturer was implying |
| $x_{59}$ | Expensive cost of living did not affect my performance in programming classes |
| $x_{60}$ | My family could afford to buy enough programming textbooks |
| $x_{61}$ | My family sponsored my academic pursuit |
| $x_{62}$ | Quarrel between family members is normal |
| $x_{63}$ | I had to travel to settle quarrels within my family |
| $x_{64}$ | Quarrel between my family members escalates a times |
| $x_{65}$ | My father is familiar with computers |
| $x_{66}$ | My mother is familiar with computers |
| $x_{67}$ | My parents are well educated |
| $x_{68}$ | My parent would want me to offer programming courses |
| $x_{69}$ | I received educational advices from family members often |
| $x_{70}$ | My family believed that a proper study will help me in programming courses |

However, twenty-one (21) factors were investigated via this study with a total of 81 variables. Each factor was coded based on the cumulative of the variables designated to investigate it as conducted by Fagbola *et al.* [11]:

*a) Student Study Habit (SSH):* This is the amount of the student's effective study in programming courses offered relative to the frequency of revision and practice and hours spent on revising the lecture notes. It was investigated by three variables $x_1$, $x_2$, $x_3$.

*b) Student Fear and Perception (SF):* This is the students' fearful perception of programming courses where a positive perception implies a reduction in fear factor of the student. This was investigated by the variables $x_4$, $x_5$, $x_6$.

*c) Student Attendance (SATD):* This is the level of effort, seriousness and devotion of students towards learning to program, investigated by the variables $x_7$, $x_8$, $x_9$.

*d) Student Attitude (SAT):* This is the level of responsiveness of a student relative to their interest, behavior and seriousness to programming courses, and characterized by student's participation in class activities, assignment, willingness to learn, and motivation from friends, colleagues and lecturer(s). This was represented by the variables $x_{10}$, $x_{11}$, $x_{12}$, $x_{13}$.

*e) Tutorials and Extra Classes (ST):* These are the extra effort put in place by students in other to have a clear

understanding of the subject matter(s) discussed programming classes. This includes extra-classes attended, assistance from friends and use of online forums and materials. This factor was investigated by the variables $x_{14}$, $x_{15}$, $x_{16}$.

*f) Lecturer Attitude (LAT):* This is defined as the lecturers' assertiveness, interest to explicitly expatiate on the subject matter, ability to motivate the student and relate with the student in a means to improve their interest in the course. This was investigated by variables $x_{17}$, $x_{18}$, $x_{19}$, $x_{20}$

*g) Teaching Style (LTS):* This is defined as the pattern of teaching of the lecturer in charge (probably dishes out voluminous handouts or excessive assignments). Whether he carries the class along and helps the student conceptualize the concept of that particular programming course. This was investigated by variables $x_{21}$, $x_{22}$, $x_{23}$, $x_{24}$.

*h) Communication Skills (LCS):* This is the ability of the lecturer to deliver the course content in a less ambiguous manner and to the understanding of the students. This entails the clarity and explicitness of the lecturer. This was investigated by variables $x_{25}$, $x_{26}$, $x_{27}$, $x_{28}$.

*i) Lecturer Availability (LA):* This is the presence and accessibility of the lecturers' when they are needed by the student(s). This factor was investigated by the variables $x_{29}$, $x_{30}$, $x_{31}$.

*j) Lecturer Dedication (LD):* This is the devotion of the lectures to the programming courses they tutor. This includes the assertiveness of the lecturers to their duty and extra effort put in place to ensure an excellent student performance. This factor was coded as presented in Table III and was investigated by the variables $x_{32}$, $x_{33}$, $x_{34}$, $x_{35}$.

*k) Health (OH):* This is the influence of medical condition on students' performance in programming courses. This factor was coded and was investigated by the variables $x_{36}$, $x_{37}$, $x_{38}$.

*l) Electricity (OE):* This is defined as the erraticism of power supply as it affects the students' practice using computers and also other laboratory works. This factor was coded and was investigated by the variables $x_{39}$, $x_{40}$, $x_{41}$.

*m)Background knowledge (OB):* This is the academic strength of the student in other courses that are elementarily related to computer programming (mathematics and physics). This factor was investigated by the variables $x_{42}$, $x_{43}$, $x_{44}$, $x_{45}$.

*n) Facilities (UF):* This is the availability of appropriate programming learning facilities (computer laboratory) within the university environment. This factor was investigated by the variables $x_{46}$, $x_{47}$, $x_{48}$, $x_{49}$ .

*o) Class population (UCP):* This is the student to tutor population ratio during the programming course class. This factor was investigated by variables $x_{50}$, $x_{51}$, $x_{52}$.

*p) Lecture time (ULT):* This is the conduciveness of the lecture schedule. This factor was investigated by the variables $x_{53}$, $x_{54}$, $x_{55}$.

*q) Teaching aids (UTA):* This is the availability of teaching aids (audio visuals) for the demonstration of the

concept of programming courses. This factor was investigated by the variables $x_{56}$, $x_{57}$, $x_{58}$.

*r) Family income (FI):* This is the robustness of the family income of the student. As it influence the ability of the student to afford textbook materials, print handout or even own a personal computer for effective study. This factor was investigated by the variables $x_{59}$, $x_{60}$, $x_{61}$.

*s) Family stress (FS):* This is the degree of disturbance from home. An unsettled home creates a paranoid atmosphere which seemly affects student performance. This factor was investigated by the variables $x_{62}$, $x_{63}$, $x_{64}$.

*t) Parent education (FPE):* This is the degree of education of the students' parent. A poor motivation from home might destabilize the student cognitive sense, hence influencing the students' performance in programming. This factor was investigated by the variables $x_{65}$, $x_{66}$, $x_{67}$.

*u) Proper guidance (FPG):* This is the student's family guidance and support level for programming courses. A student from a family of computer scientist is prone to having huge support and guidance from home. This factor was investigated by the variables $x_{68}$, $x_{69}$, $x_{70}$.

After final normalization and cleaning process were completed, the entire data acquired was divided into variable and factor datasets and each data split was used to train the machine learning classifiers.

*C. Development of the Machine learning-based Student Performance Predictive Models*

M5P decision tree and the linear regression classifier, having industrially-packaged working implementations in WEKA environment, were trained using the variable and factor datasets and further applied to generate predictive models which are of exclusive significance to the determination of students' performance. The variable-based student performance model generated by the linear regression classifier is presented in (5).

$$
\begin{aligned}
x80 &= 0.0444 * x1 + 0.3166 * x2 + 0.0746 * x3 - 0.0415 * \\
&x4 - 0.239 * x5 + 0.3153 * x6 - 0.1467 * x7 + 0.3464 * \\
&x8 + 0.6227 * x9 - 0.1404 * x11 - 0.3228 * x12 + \\
&0.1179 * x13 - 0.4613 * x14 - 0.3948 * x15 + 0.4249 * \\
&x16 - 0.2241 * x17 - 0.1389 * x18 + 0.2025 * x19 + \\
&0.0664 * x20 + 0.133 * x21 + 0.1745 * x22 - 0.3222 * \\
&x23 - 0.3334 * x24 - 0.2479 * x25 - 0.1623 * x26 + \\
&0.0665 * x28 - 0.2556 * x29 + 0.2829 * x30 - 0.2215 * \\
&x31 - 0.4575 * x33 + 0.135 * x34 + 0.3312 * x35 - \\
&0.2152 * x36 + 0.2407 * x37 + 0.1757 * x38 - 0.2986 * \\
&x39 + 0.1768 * x40 - 0.2375 * x41 - 0.1969 * x42 + \\
&0.2352 * x43 - 0.098 * x44 + 0.4561 * x45 - 0.136 * \\
&x46 - 0.387 * x47 + 0.1525 * x48 - 0.2215 * x49 + \\
&0.0481 * x50 + 0.1292 * x51 + 0.1508 * x52 + 0.4368 * \\
&x53 - 0.3313 * x54 - 0.1794 * x55 - 0.0523 * x56 - \\
&0.3505 * x57 + 0.4718 * x58 + 0.269 * x59 + 0.086 * \\
&x60 - 0.3004 * x61 - 0.444 * x62 + 0.3544 * x63 - \\
&0.2301 * x64 - 0.538 * x65 + 0.0899 * x66 + 0.2394 * \\
&x67 - 0.0681 * x68 - 0.1007 * x69 - 0.3858 * x70 + \\
&9.8865
\end{aligned}
\tag{5}
$$

The learned models developed are further used to generate predictions on new instances. The factor-based Student

Performance Model obtained using linear regression classifier is expressed in (6).

$$grade = -0.074 * sf + 0.0942 * satd + 0.065 * sat + 0.0449 * lat - 0.0448 * lcs - 0.0407 * la + 0.0493 * oh + 0.0814 * oe - 0.0792 * uf + 0.0621 * fi - 0.0663 * fs - 0.0533 * fpe - 0.1233 * fpg + 5.6703 \qquad (6)$$

The M5 pruned model tree for the variable dataset is presented in Fig. 1. However, the variable-based M5P decision tree classifier generated smoothed Linear Models (LM) through 22 refinement processes. The first and the last generated models are presented in (7) and (8), respectively although the latest refinement was used to predict student performance.

$$x80 = 0.0297 * x1 + 0.0187 * x4 - 0.0376 * x5 + 0.1263 * x9 - 0.017 * x12 - 0.0826 * x14 + 0.021 * x15 + 0.0316 * x19 - 0.0209 * x22 + 0.0389 * x57 + 0.0211 * x59 - 0.0343 * x65 - 0.0217 * x69 + 3.9539 \qquad (7)$$

$$x80 = 0.0155 * x1 + 0.0098 * x4 - 0.0652 * x5 + 0.0552 * x7 + 0.1046 * x9 - 0.0089 * x12 - 0.0143 * x14 + 0.011 * x15 - 0.0503 * x19 - 0.1112 * x22 + 0.032 * x29 - 0.0288 * x33 + 0.0324 * x59 - 0.06 * x65 - 0.188 * x69 + 5.9906 \qquad (8)$$

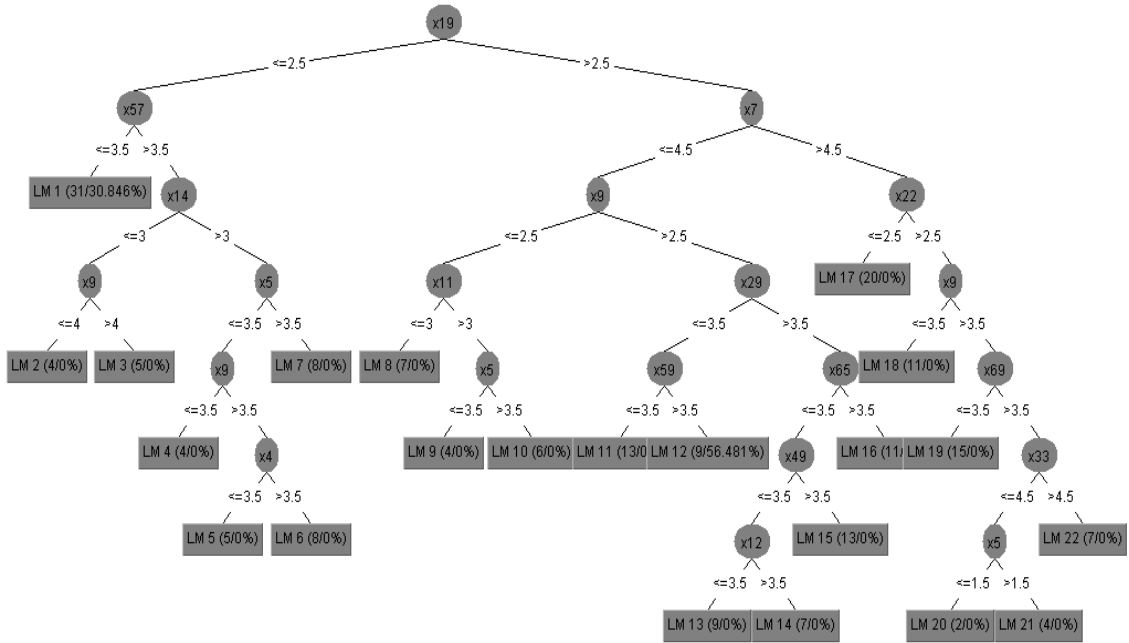

Fig. 1. The M5 pruned model tree for the variable dataset.

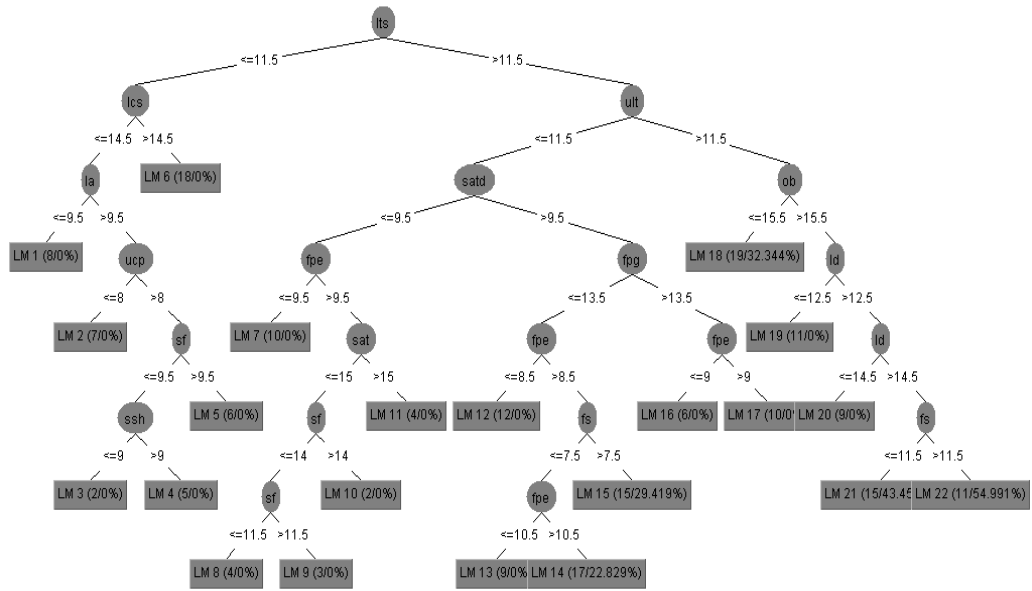

Fig. 2. The M5 pruned model tree for the factor dataset.

The M5 Pruned model tree for the factor dataset is presented in Fig. 2. However, the factor-based M5P classifier generated smoothed Linear Models (LM) through 22 refinement processes. The first and the last models generated are presented in (9) and (10), respectively.

$grade = 0.0481 * ssh + 0.1057 * sf + 0.0343 * satd + 0.0084 * sat - 0.0083 * st + 0.0127 * lat + 0.0475 * lcs - 0.1963 * la + 0.0141 * oe + 0.0232 * ucp - 0.0248 * fs - 0.0097 * fpe - 0.0293 * fpg + 5.0805$ (9)

$grade = -0.0144 * sf + 0.0307 * satd + 0.0106 * sat - 0.0101 * st + 0.0045 * lat - 0.0548 * lcs + 0.0202 * ld + 0.0273 * oe + 0.0675 * ob - 0.0159 * ult - 0.0466 * fs - 0.0034 * fpe - 0.0321 * fpg + 4.0297$ (10)

### D. Validation of the Developed Machine Learning-based Student Performance Predictive Models

The variable and factor datasets were employed in the development of the students' performance predictive models, which were then validated using the test dataset. Some instances of the validation results of the predictive models generated by the machine learning classifiers are presented in Table II. It is important to note that with limited data used for validation, the results of validation test cannot be exclusively used to justify the correctness of the developed models but rather by some standard evaluation measures. Based on some validation results obtained, the best performing model is the factor dataset-based SPM generated by the linear regression classifier. This is followed by variable dataset-based SPM generate by M5P decision tree classifier, factor dataset-based M5P decision tree and the variable dataset-based SPM based on linear regression classifier in decreasing order of performance. Note that the best prediction values are marked in "bold".

TABLE II. MODEL VALIDATION INSTANCES FOR LINEAR REGRESSION AND M5P DECISION TREE CLASSIFIERS

| Actual Grade | Linear Regression Classifier Algorithm (variable dataset)-based SPM | Linear Regression Classifier Algorithm (factor dataset)-based SPM | M5P Decision Tree Classifier (variable dataset)-based SPM | M5P Decision Tree Classifier (factor dataset)-based SPM |
|---|---|---|---|---|
| 4 | 4.3618 | **4.0124** | 3.865004 | 4.0347 |
| 6 | 6.2135 | 5.9675 | **5.96883** | 5.9505 |
| 4 | 4.2946 | 4.1055 | **4.036288** | 4.1578 |
| 6 | 5.0878 | **5.9583** | 5.375602 | 5.2558 |
| 5 | 4.6443 | **5.0572** | 4.774742 | 4.4751 |
| 5 | 5.1855 | **4.9381** | 4.881071 | 5.2582 |
| 6 | 5.3058 | **5.8879** | 6.184321 | 5.8878 |
| 5 | **4.8282** | 4.8246 | 4.146855 | 4.8255 |
| 6 | 5.7855 | 6.5766 | 5.697118 | **5.9423** |
| 6 | 4.3962 | **6.039** | 5.271766 | 5.3175 |

### E. Performance Evaluation Metrics for the Machine Learning-based Approaches Used

The mean absolute error, root mean square error, relative absolute error, root relative squared error, time taken to build and test the models are the standard metrics used to evaluate the performance of the learning techniques.

*a) Root Relative Squared Error (RRSE)* is determined using the relation:

$$RRSE = \sqrt{\frac{\sum_{j=1}^{n}(P_{(ij)} - T_j)^2}{\sum_{j=1}^{n}(T_j - \bar{T})^2}}$$ (11)

where $P_{(ij)}$ represents the predicted value by each individual program $i$ for any sample case $j$ which is a subset of $n$ sample cases, $T_j$ is the target value for sample case $j$; and $\bar{T}$ is given by [28]:

$$\bar{T} = \frac{1}{n}\sum_{j=1}^{n}T_j$$ (12)

*b) The Relative Absolute Error, RAE,* accepts the total absolute error and divides it with the actual absolute error of the model predictor. Relative Absolute Error is determined using the relation [24]:

$$RAE = \frac{\sum_{j=1}^{n}|P_{(ij)} - T_j|}{\sum_{j=1}^{n}|T_j - \bar{T}|}$$ (13)

*c) Mean Absolute Error, MAE,* is determined by adding the absolute values of the error, $e_i$, and then dividing the total error by $n$ [24]:

$$MAE = \frac{1}{n}\sum_{i=1}^{n}|e_i|$$ (14)

*d) Root Mean Square Error:* This is a measure of the differences between the sample values predicted by a model and those which are actually observed from the system that is being modelled [28]. That is, the change between the model performance of a predictive model and another. Analytically,

$$RMSE = \sqrt{MSE}$$ (15)

where $MSE = \sum_{i=1}^{n}\frac{(\mu_i - \hat{z}_i)^2}{n}$ such that $\hat{z}_i$ is the model-predicted response for input $x_i$.

*e) Time taken to build the model*: This is the total time required to learn the discriminating features and to develop a model

*f) Time taken to test the model:* This is the time taken to validate and ascertain the correctness of the developed model.

### III. THE DESIGN AND IMPLEMENTATION OF A MOBILE FRONT-END APPLICATION FOR THE DEVELOPED PREDICTIVE MODELS

The developed student performance models were implemented within an Android 1.0.1 Studio environment, using XML for the design of the Graphical User Interface (GUI) and Java for the logic that unifies the GUI and the implementation of the developed models. The flowchart representation for the implementation of the developed student performance models is presented in Fig. 3. The code and design interface is presented in Fig. 4. In the same vein, the mobile home interface of the SPM implementation as presented in Fig. 5 defines the model(s) to be applied and serves as a link to the questioning aspects of the application. Students and stakeholders can predict the performance of a student by selecting any of the options presented on the home activity of the application. Each of these options implement an underlying model which is used for the prediction of student performance relative to their responses to questions presented.

The interface presented in Fig. 6 displays various questions which are relevant to the selected prediction perspectives. Responses to these questions are then interlinked with the underlying models. In Fig. 7, the predicted performance of the student is displayed in an alert message-box after the responses from prospective students and educational stakeholders have been substituted into the chosen model(s). This happens upon clicking the finish button which appears after the entire questions required for the prediction of student performance under the selected perspective has been duly responded to.

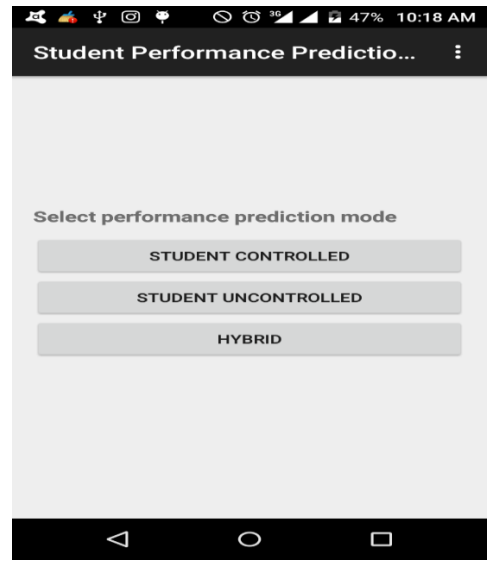

Fig. 5.    Home interface of the mobile student performance evaluator.

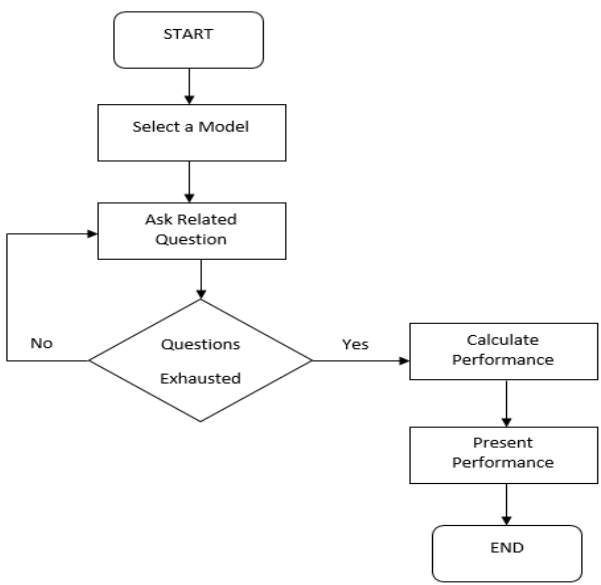

Fig. 3.    Flow control of the implementation of student performance models.

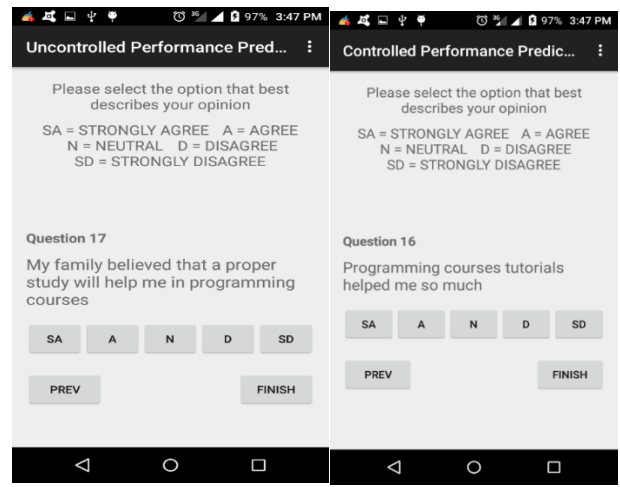

Fig. 6.    Interface of the implemented SP models.

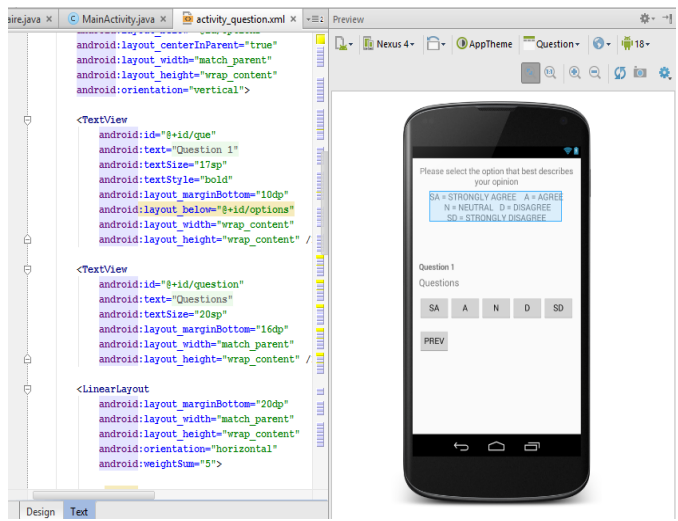

Fig. 4.    Code and design interface of the student performance models.

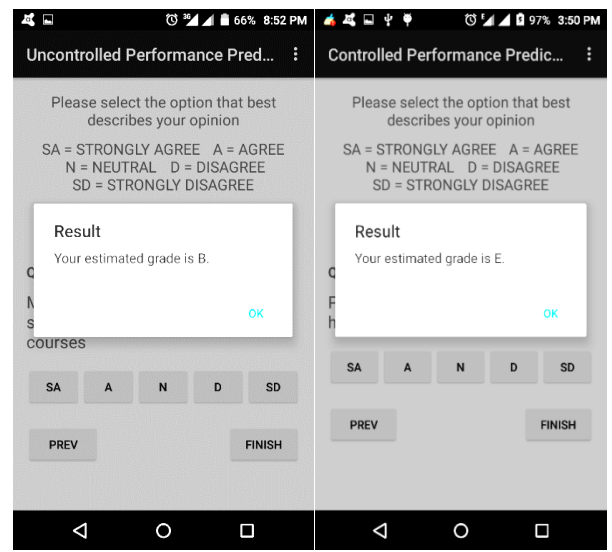

Fig. 7.    Instances of predicted students' performance.

## IV. RESULTS AND DISCUSSION

In this section, the performance and comparative evaluation results of the machine-learning predictive approaches and the developed student performance models are presented and discussed.

### A. Results of Performance Evaluation of the Machine Learning Methods

The results regarding the mean absolute error, root mean square error, relative absolute error, root relative squared error, time taken to build and test the models for both linear regression and M5P decision tree classifiers are presented in Table III. The variable-based Linear Regression Classifier produced the best model in terms of mean absolute error, root mean squared error, relative absolute error and the root relative squared error having yielded the least values in all these metrics. This is followed by the variable-based M5P decision tree, factor-based M5P Decision Tree and the factor-based linear regression classifiers in decreasing order of performance. In terms of the time to build the model, the results obtained indicate that the factor-based M5P Decision Tree is the most computationally-efficient classifier followed by variable-based Linear Regression classifier, variable-based M5P decision tree and factor-based linear regression classifier.

TABLE III. MODEL VALIDATION INSTANCES FOR LINEAR REGRESSION AND M5P DECISION TREE CLASSIFIERS

| Technique | Mean Absolute Error | Root Mean Square Error | Relative Absolute Error (%) | Root Relative Squared Error (%) | Time taken to build model (s) | Time taken to test model (s) |
|---|---|---|---|---|---|---|
| Linear Regression Classifier (variable-based) | 0.1638 | 0.2386 | 20.307 | 24.8369 | 0.09 | 0.03 |
| Linear Regression Classifier (factor-based) | 0.5853 | 0.7273 | 72.5498 | 75.7246 | 0.25 | 0.06 |
| M5P Decision Tree Classifier (Variable-based) | 0.3054 | 0.4067 | 41.0867 | 47.2537 | 0.13 | 0.02 |
| M5P Decision Tree Classifier (Factor-based) | 0.3984 | 0.555 | 53.6099 | 64.4848 | 0.05 | 0.01 |

Using the model produced by the best performing classifier (variable-based LRC), three (3) out of the 70 variables investigated are found to be insignificant to student performance as presented in Table IV. However, there are 32 variables with positive significance and 35 variables with negative significance to student performance in programming courses as presented in Tables V and VI, respectively.

TABLE IV. VARIABLE-BASED LRC' SPM VARIABLES WITH INSIGNIFICANT EXPRESSIONS

| S/N | Insignificant Expressions |
|---|---|
| $x_{10}$ | I believed I could understand the programming course |
| $x_{27}$ | Programming language lecturers delivered course contents well and to my understanding |
| $x_{32}$ | Programming course lecturers allowed students to ask questions and take time to explain |

TABLE V. VARIABLE-BASED LRC' SPM VARIABLES WITH POSITIVE EXPRESSIONS

| S/N | Expressions with Positive Significance |
|---|---|
| $x_1$ | I had enough time to study programming |
| $x_2$ | Studying before attending a class aided my assimilation during programming classes. |
| $x_3$ | Studying programming was never a wasted effort |
| $x_6$ | I was always nervous during programming examinations |
| $x_8$ | Blending in after missing a class was very easy |
| $x_9$ | I was very serious with programming classes |
| $x_{13}$ | Programming is relevant to my pursuit |
| $x_{16}$ | Programming courses' tutorials helped me so much |
| $x_{19}$ | Programming courses' lecturers were never partial in their dealings with students |
| $x_{20}$ | Programming courses' lecturers were friendly during lectures |
| $x_{21}$ | Programming courses' lecturers enforced discipline during their lectures |
| $x_{22}$ | Programming courses' lecturers were too serious during lectures |
| $x_{28}$ | Programming courses' lecturers were very clear and explicit |
| $x_{30}$ | Programming courses' lecturers attended to me whenever I had difficulties with their course(s) |
| $x_{34}$ | Programming courses' lecturers spent extra time to explain things during class |
| $x_{35}$ | Programming courses' lecturers usually came early to class |
| $x_{37}$ | Prolong usage of computer caused me headache |
| $x_{38}$ | I took a few compulsory medications frequently |
| $x_{40}$ | Erratic power supply reduced the effectiveness of my practice |
| $x_{43}$ | I had a good background in mathematics |
| $x_{45}$ | Strong background in Physics and Mathematics helped me in programming |
| $x_{48}$ | Lack of computer programming facilities disrupted clear understanding of programming lessons |
| $x_{50}$ | Large class population disrupted my concentration during programming lectures |
| $x_{51}$ | Population of students offering programming courses debarred my commitment to learning |
| $x_{52}$ | Effectiveness of the programming lecturers' teaching was reduced by huge programming class population. |
| $x_{53}$ | Programming lectures were scheduled after an equally tiring lecture |
| $x_{58}$ | I had a visual understanding of what the programming lecturer was implying |
| $x_{59}$ | Expensive cost of living did not affect my performance in programming classes |
| $x_{60}$ | My family could afford to buy enough programming textbooks |
| $x_{63}$ | I had to travel to settle quarrels within my family |
| $x_{66}$ | My mother is familiar with computers |
| $x_{67}$ | My parents are well educated |

TABLE VI.     VARIABLE-BASED LRC' SPM VARIABLES WITH NEGATIVE EXPRESSIONS

| S/N | Expressions with Negative Significance |
|---|---|
| $x_4$ | Programming sounded very scary |
| $x_5$ | I was always nervous during programming classes |
| $x_7$ | I attended programming classes regularly |
| $x_{11}$ | I had interest in programming beyond class level |
| $x_{12}$ | Programming was not confusing and did not cause headache |
| $x_{14}$ | Group discussions helped me to understand programming |
| $x_{15}$ | Attending programming tutorials was very helpful |
| $x_{17}$ | Motivation of programming lecturers encouraged my commitment towards learning programming |
| $x_{18}$ | Programming language lecturers helped me develop interest in programming |
| $x_{23}$ | Teaching methods and styles of programming lecturers inhibited lecture clarity |
| $x_{24}$ | Programming language lecturers wasted time on matters with less relevance in class |
| $x_{25}$ | Programming language lecturers were always clear, precise and communicates understandably |
| $x_{26}$ | Programming language lecturers made use of enough relevant instructional materials |
| $x_{29}$ | Programming language lecturers didn't miss classes |
| $x_{31}$ | Programming lecturers were always available |
| $x_{33}$ | Programming course lecturers came to class fully prepared |
| $x_{36}$ | I fell sick quite often |
| $x_{39}$ | It was difficult to charge my computer even within the campus |
| $x_{41}$ | Consistent power supply helped me in programming courses |
| $x_{42}$ | I had a good background in physics |
| $x_{44}$ | I had a good background in English |
| $x_{46}$ | Absence of accessible ICT facilities inhibited my programming performance |
| $x_{47}$ | The environment where we had programming lectures was not conducive |
| $x_{49}$ | The school library was not equipped with materials relevant to programming |
| $x_{54}$ | Programming courses were scheduled to non-conducive times |
| $x_{55}$ | We had programming classes at unfavorable times |
| $x_{56}$ | Programming lecture theatres were equipped with audio-visuals and learning aids |
| $x_{57}$ | Programming courses were analyzed clearly to sight |
| $x_{61}$ | My family sponsored my academic pursuit |
| $x_{62}$ | Quarrel between family members is normal |
| $x_{64}$ | Quarrel between my family members escalates a times |
| $x_{65}$ | My father is familiar with computers |
| $x_{68}$ | My parent would want me to offer programming courses |
| $x_{69}$ | I received educational advices from family members often |
| $x_{70}$ | My family believed that a proper study will help me in programming courses |

### B. Comparative Evaluation of the Developed Student Performance Models

The expressions of variable-based LRC model with positive significance agree with some already established variables such as students' lack of understanding, absence from class, negative attitudes towards programming, students' performance in Mathematics [29], study habit [30], review study materials, self-evaluate, rehears explaining materials, and studying in a conducive environment [31], students' class attendance (Pudaruth, Nagowah, Sungkur, Moloo and Chinia [32], Teaching Styles and Strategies [33], availability of University facilities [6] and mathematics background [34]. However, this study established the negative significance of variables such as group discussions, good background in physics and English among others on student performance in programming as against the reports of Mohd and Abdullah

[29] and Darwin *et al.* [30] for example. In general, the variable-based LRC model is an explicit extension of most existing counterparts by salient factors such as Lecturers' Teaching Style (LTS), Health (OH), Electricity (OE), Parental Education (FPE), Student Fear and Perception (SF), Tutorials and Extra Classes (ST) among others which have not been duly considered by other previous works.

## V.     CONCLUSION AND FUTURE WORKS

This study was conducted to explore the factors affecting the academic performance of undergraduates in programming courses and develop models with which the performance of students can be predicted. The research was conducted on a sample of students who have at one time or the other offered Web programming, C or JAVA within the Federal University, Oye-Ekiti, Ekiti State, Nigeria between 2012 and 2016. This was based on students' performance records which cut across the second and third (200-300) levels of study within the institution. Machine learning approaches were gainfully employed for the analysis of the retrieved data from a defined number of respondents. Results obtained indicate that the attitude of students and lecturers, fearful perception of students, erratic power supply, university facilities, student health, students' attendance are significant to the performance of students in programming courses. It is recommended that future research adopts improved statistical machine learning approaches to comparatively model the learning behaviour in private and public Universities of Nigeria and identify the salient factors significant to performance of students in both systems for robust evaluation of quality of training and to aid effective decision making by the government, students and University education stakeholders. Furthermore, a consideration of all programming courses being offered in the institution and a relatively larger population might graciously improve the findings reported in this study. The existing statistical machine learning approaches can also be extended while some other ones can be introduced for more accurate results.